\documentclass{optica-article}
\articletype{Research Article}

\usepackage{lineno}
\usepackage{amsmath,amsfonts}
\usepackage{graphicx}
\usepackage{booktabs}
\usepackage{textcomp}

\begin{document}

\title{110-GHz bandwidth integrated lithium niobate modulator without direct lithium niobate etching}

\author{Yifan Qi\authormark{1}, Gongcheng Yue\authormark{1}, Ting Hao\authormark{2} and Yang Li\authormark{1,*} }

\address{\authormark{1}State Key Laboratory of Precision Measurement Technology and Instruments, Department of Precision Instrument, Tsinghua University, Beijing, 100084, China\\

\authormark{2}Advanced Fiber Resources (Zhuhai), Ltd., Zhuhai, 519080, China\\}

\email{\authormark{*}yli9003@tsinghua.edu.cn} 


\begin{abstract*} 
Integrated thin film lithium niobate (TFLN) modulators are emerging as an appealing solution to high-speed data processing and transmission due to their high modulation speed and low driving voltage. The key step in fabricating integrated TFLN modulators is the high-quality etching of TFLN, which typically requires long-term optimization of fabrication recipe and specialized equipment. Here we present an integrated TFLN modulator by incorporating low-index rib loaded waveguides onto TFLN without direct etching of TFLN. Based on our systematic investigation into the theory and design methodology of the proposed design, we experimentally demonstrated a TFLN etching-free Mach-Zehnder modulator, featuring a flat electro-optic response up to 110 GHz and a voltage-length product of 2.53 V cm. By significantly simplifying the fabrication process, our design opens up new ways of mass production of high-speed integrated TFLN modulators at low cost.

\end{abstract*}

\section{Introduction}
High-speed integrated lithium niobate modulators are emerging as an promising solution for applications in fiber-optic communications, microwave photonics, and quantum optics\cite{chen2022advances,lin2020advances, qi2020integrated, boes2023lithium}. Distinct from silicon photonic modulators, thin film lithium niobate (TFLN) modulators take advantage of the intrinsic electro-optic effect to achieve a high modulation speed, a low driving voltage, and a low insertion loss\cite{zhu2021integrated} without doping in the fabrication process. To date, most high-performance TFLN modulators require the nanofabrication of TFLN based on argon plasma etching process. Although this dry etching process can yield TFLN ridge waveguides with a propagation loss as low as 3 dB/m, this dry etching process still shows a relatively low selectivity and a low reproduction rate in different dry etching equipments.\cite{he2019high, wang2018integrated, kharel2021breaking, hu2021folded, xu2020high, li2020lithium, xu2022dual, luke2020wafer,lin2022high}. Moreover, lithium is regarded as a contaminating element of CMOS process, necessitating the specialized etching equipment and in turn dramatically increasing the fabrication cost. Dry etching-based nanofabrication recipe's low reproduction rate and high cost hinder the mass production of high-performance integrated TFLN modulators at low cost.

Direct etching of TFLN can be avoided in the fabrication process of integrated TFLN modulators based on the rib loaded waveguide structure\cite{hunsperger1995integrated}. In this structure, a rib structure is fabricated on top of the TFLN, leading to an increase in the effective refractive index underneath the rib and the confinement of the optical mode\cite{rao2017compact, hunsperger1995integrated}. The rib is typically fabricated by tantalum pentoxide (Ta$_2$O$_5$) \cite{rabiei2013heterogeneous}, titanium dioxide (TiO\(_2\)) \cite{li2015waveguides}, and silicon nitride (Si\(_3\)N\(_4\)) \cite{jin2015linbo, rao2016high, ahmed2020subvolt, ahmed2020high, boynton2020heterogeneously, huang2021high, zhang2021high, nelan2022ultra} which are much easier to deposit and pattern than lithium niobate. However, because these materials show similar refractive indices to that of lithium niobate, a large portion of the light mode is confined in the rib without electro-optic effect, decreasing electro-optic coupling efficiency.

By introducing bound state in the continuum (BIC) to rib loaded waveguide structures, the light mode can be strongly confined in the TFLN via a lower index rib \cite{yu2019photonic, yu2020high}. However, to achieve the BIC mode with a low optical loss, these structures can only show transverse magnetic (TM) bound state in the transverse electric (TE) continuum \cite{zou2015guiding}, limiting the polarization state of the TFLN modulator to TM mode. For such a modulator, to take advantage of the maximum electro-optic coefficient $r_{33}$ ($\sim31$ pm/V) of lithium niobate, a \textit{z}-cut TFLN with out-of-plane electrodes should be used, leading to a complicated electrode structure and a challenging fabrication process. Alternatively, if we choose an \textit{x}-cut TFLN with in-plane electrodes, we can only modulate the TM mode via a small electro-optic coefficient $r_{31}$ of TFLN ($\sim10$ pm/V), leading to a high driving voltage.

Here we demonstrate a high-speed silica rib loaded waveguide TFLN modulator for tight confinement of optical TE mode. We systematically analyzed the theory and design methodology of TFLN rib loaded waveguide modulator. Based on this analysis, we selected silica as the rib material and carefully designed the geometric parameters of this modulator, leading to a good confinement of TE mode in the TFLN area underneath the silica rib. We fabricated and tested this design, featuring a flat electro-optic response above -3 dB of over 110 GHz and a half-wave voltage-length product of 2.53 V cm. While showing a performance comparable to that of etched TFLN ridge waveguide modulators, our design significantly simplifies the fabrication process by eliminating the dry etching of TFLN, leading to a fabrication recipe with high reproduction rate and low cost. Compared with BIC-based rib loaded waveguide TFLN modulators, our design enables modulating TE mode via the highest electro-optic coefficient $r_{33}$ of TFLN, resulting in a low driving voltage and a simple in-plane electrode structure. By achieving high-performance TFLN modulator without direct TFLN etching, our design provides a promising solution to the mass production of high-performance integrated lithium niobate modulators at low cost.

\section{Theory and design methodology of low index rib loaded waveguide}

\begin{figure}[ht!]
\centering\includegraphics[width=\linewidth]{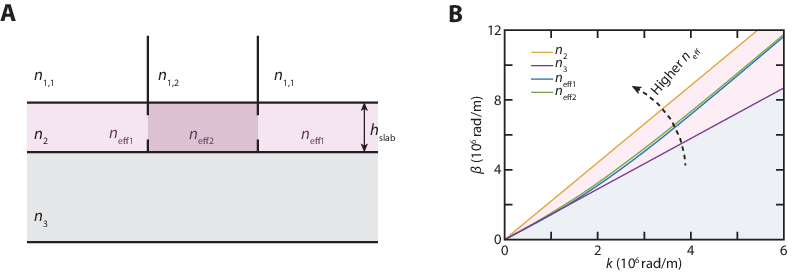}
\caption{Schematic and dispersion relationships of the fundamental TE mode of low-index rib loaded waveguide structure. (A) Diagram of a rectangular rib loaded waveguide. (B) Dispersion relationships of different regions of the structure in (A). Pink region represents the guided mode while gray region represents the leaky mode.}
\label{Structure}
\end{figure}

We first analyze the rib loaded waveguide from the viewpoint of the dispersion relationship of fundamental TE mode. Based on the TE mode of rib loaded waveguide, we can take advantage of the highest electro-optic coefficient $r_{33}$ of lithium niobate with in-plane traveling wave electrode. The rib loaded waveguide structure can be equivalently simplified as the structure in figure \ref{Structure}A. In this structure, the slab with refractive index $n_2$ can be divided into three regions: the rib loaded waveguide region underneath the rib (with effective index $n_{\rm {eff2}}$) and two adjacent regions (with effective index $n_{\rm {eff1}}$). The TE-mode effective indices of these regions can be approximately calculated by the planar waveguide theory \cite{hunsperger1995integrated}, leading to the dispersion equation:
\begin{equation}
 \tan(a h_{\rm slab})=\frac{a(p+q)}{a^2-pq}
\end{equation} where \( a=\sqrt{n_2^2 k^2-\beta^2}\), \( p=\sqrt{\beta^2-n_1^2 k^2}\), \( q=\sqrt{\beta^2-n_3^2 k^2}\), \(h_{\rm slab}\) is the slab thickness, \( \beta \) is the propagation constant of the mode, and \( k \) is the propagation constant of free space\cite{hunsperger1995integrated}. Based on the material and geometric properties of rib loaded waveguide structure consisting of air superstrate, silica rib, TFLN slab, and silica substrate, we calculate the dispersion relationships of rib loaded waveguide region and the adjacent regions (figure\ref{Structure}B) for fundamental TE mode. By setting the rib with a refractive index $n_{\rm {1,2}}$ higher than that of the ambient medium $n_{\rm {1,1}}$, the rib loaded waveguide region can always show an effective index $n_{\rm {eff2}}$ higher than that of the adjacent regions $n_{\rm {eff1}}$ even when $n_{\rm {1,2}}$ is not close to $n_{\rm {2}}$. This conclusion brings more freedom in selecting the rib material for TE mode, leading to a greater potential in improving the performance of integrated rib loaded waveguide modulator.

To fully take advantage of the highest electro-optic coefficient $r_{33}$ of x-cut TFLN to achieve an effective electro-optic modulation, the fundamental TE mode should be confined in the TFLN slab rather than the rib. Moreover, to minimize the propagation loss of the fundamental TE mode, the Ohmic loss induced by the metal electrodes should be reduced as much as possible. The confinement of fundamental TE mode in the slab and propagation loss of light are effected by the rib material, height and width of the rib, as well as thickness of the TFLN slab. Here, we quantitatively analyze these effects by using numerical simulations.

\begin{figure}[ht!]
\centering\includegraphics[width=\linewidth]{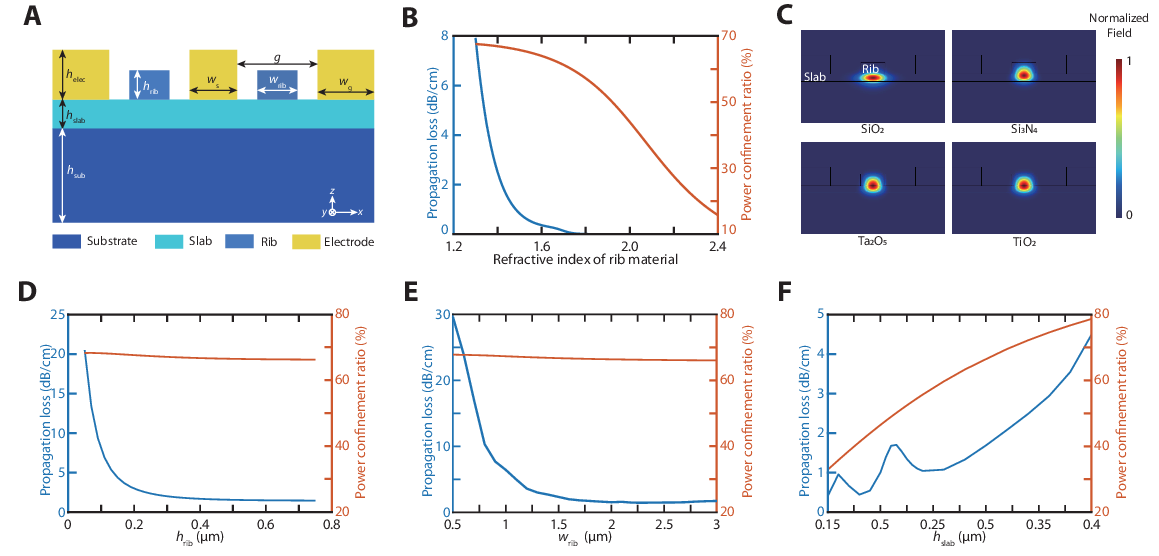}
\caption{Numerical simulations of TFLN rib loaded waveguide structure for electro-optic modulator. (A) Cross-section view of TFLN Mach-Zender electro-optic modulator based on rib loaded waveguide structure. (B) Propagation loss and power confinement ratio as a function of the refractive index of rib. (C) Optical modes of rib loaded waveguide modulators with different rib materials and the same geometric parameters. The material of the rib is shown below each figure. All the modulators are with TFLN slab, silica substrate, and gold electrodes. (D)-(F) Propagation losses and power confinement ratios as a function of (D) silica rib height, (E) silica rib width, and (F) thickness of the slab. The power confinement ratio is defined as the ratio of the optical power confined in the slab to the total optical power.}
\label{Simulation}
\end{figure}

We first analyze the effect of the rib's refractive index on the performance of rib loaded waveguide structure for electro-optic modulator by using Lumerical MODE (figure \ref{Simulation}A). As shown in figure \ref{Simulation}B, as the refractive index of the rib (\(n_{\rm rib}\)) decreases, the power confinement ratio of fundamental TE mode inside the TFLN slab increases, resulting in a higher electro-optic modulation efficiency. As \(n_{\rm rib}\) further decreases, the effective index of the mode decreases, leading to a weaker mode confinement in the transverse direction. Such a weaker mode confinement results in a larger bending radius and a higher propagation loss induced by the metal electrodes. To better visualize the mode profiles of rib loaded waveguide-based modulators with different rib indices, we simulated the optical modes of rib loaded waveguide-based modulators with rib fabricated by silica, silicon nitride, tantalum pentoxide, and titanium dioxide. As shown in figure \ref{Simulation}C, compared with high-index ribs, a low-index rib can confine a larger portion of the mode in the TFLN slab, leading to a higher electro-optic modulation efficiency and a lower half-wave voltage. However, a low-index rib results in a weaker lateral confinement, leading to a higher Ohmic loss induced by the metal electrodes.

We then study the effect of the geometric parameters of TFLN slab and silica rib on the performance of rib loaded waveguide for electro-optic modulator. As shown in figure \ref{Simulation}D, E, as silica rib height (\(h_{\rm rib}\)) and width (\(w_{\rm rib}\)) increase, we can achieve a lower propagation loss while maintaining a high mode confinement. However, a very large \(h_{\rm rib}\) will increase the fabrication difficulty. And, a very large \(w_{\rm rib}\) will result in a high Ohmic loss induced by the large mode area and the metal electrodes, as well as more high-order modes. Following these rules, we could design (\(h_{\rm rib}\)) and (\(w_{\rm rib}\)) with reasonable values. As shown in figure \ref{Simulation}F, as TFLN slab height increases, the lateral confinement of the fundamental TE mode becomes weaker, leading to a higher propagation loss induced by the metal electrodes. On the other hand, a very small \( h_{\rm slab}\) will result in a smaller portion of mode in the TFLN slab due to the strong leakage to the silica substrate, leading to a lower electro-optic modulation efficiency and a higher half-wave voltage. Hence, we should choose a reasonable value of \( h_{\rm slab}\) to maintain a good balance between propagation loss and modulation efficiency.

Based on the analysis of the effects of rib index, geometric parameters of rib and TFLN slab on optical mode, we can design a better rib loaded waveguide-based electro-optic modulator. Such a modulator with the rib index lower than that of the TFLN slab can maintain a strong confinement of fundamental TE mode in the slab under various geometric parameters, leading to a low half-wave voltage. In the design of a rib loaded waveguide-based electro-optic modulator, the geometric parameters such as width and height of the rib as well as thickness of the slab are determined not only by the waveguide design, but also by the electrodes design considering impedance and velocity matching. This electrodes design will have more degrees of freedom in the consideration of our analysis of the waveguide configuration.

\section{Design, fabrication, and measurement of low-index rib loaded waveguide modulator}

We designed a silica rib loaded waveguide-based Mach-Zender modulator to verify the performance of low-index rib loaded waveguide-based modulator. We used traveling wave electrodes to achieve a better high-speed modulation performance. These traveling wave electrodes can be designed by tuning the geometric parameters (\(h_{\rm rib}\) and \(w_{\rm rib}\) in figure \ref{Simulation}A) over a huge range due to the optical mode's great robustness against these parameters' variations (figures \ref{Simulation}D, E). Here we adjust the geometric parameters by simultaneously optimizing the optical and electric figures of merit, leading to a good high-speed modulation performance and low fabrication difficulty. The optimized geometric parameters are: \(w_{\rm rib} = 2\) \textmu m, \(h_{\rm rib}=400\) nm, \(h_{\rm elec}=800\) nm, \(h_{\rm slab}=300\) nm, and \(h_{\rm sub}\)=3 um. To improve the adhesion to TFLN, the electrodes are consisting of 100-nm titanium and 700-nm gold along \( +z \) (figure \ref{Simulation}A). To control the optical loss, we designed \(g\) as 7 \textmu m. To easily feed the microwave signal to electrodes via microwave probes, we designed \(w_s\) as 20 \textmu m and \(w_g\) as 150 \textmu m.

\begin{figure}[ht!]
\centering\includegraphics[width=\linewidth]{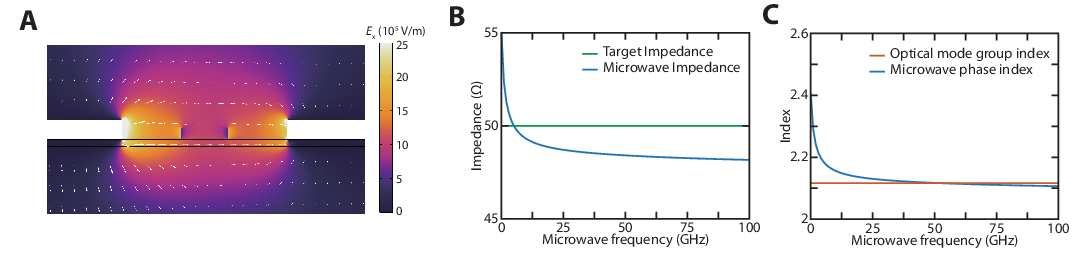}
\caption{Electric simulation results of silica rib loaded waveguide-based electro-optic modulator. (A) Electric field distribution over the waveguide region. Arrows show the direction of electric field. (B) Microwave impedance matching and (C) velocity matching results of the electro-optic modulator with traveling wave electrodes.}
\label{Electric}
\end{figure}

Compared with etched TFLN ridge waveguide, silica rib loaded waveguide also enhances the modulation electric field within the TFLN waveguide. The modulation electric field couples into etched TFLN ridge waveguide along the direction perpendicular to the sidewall of the etched waveguide, leading to low modulation electric field within the etched waveguide due to its high index contrast at microwave frequencies. In contrast, the modulation electric field couples into TFLN underneath the rib mainly along the direction parallel to the TFLN top surface (arrows in figure \ref{Electric}A), leading to a high modulation electric field in the TFLN region underneath the rib. In this region, the average electric field intensity is as high as \(1.115 \times 10^5 \) V/m. As shown in figures \ref{Electric} B and C, the optimized design can achieve good impedance matching and group velocity matching when the modulation frequency varies from DC to 100 GHz. We could further improve the high-speed modulation performance by targeting at a higher modulation frequency with good impedance matching and group-velocity matching.

Our design of silica rib loaded waveguide-based modulator was fabricated using a standard planar process based on a TFLN wafer. To achieve the geometric parameters of our design, the TFLN wafer was customized with following parameters: 300-nm-thick \(x\)-cut TFLN, 3-\textmu m-thick silica buried layer, and 525-\textmu m-thick silicon substrate. To fabricate the rib, a 400-nm-thick silica film was first deposited on top of the wafer by plasma enhanced chemical vapor deposition. The waveguide pattern was then defined with electron beam lithography (EBL) by negative EBL resist. Then the silica film was fully etched with a standard reactive ion etching process. Because silica etching is more well-developed than lithium niobate etching, this step dramatically simplifies the entire fabrication process, leading to a higher reproduction rate. Finally, the electrodes were fabricated with a standard lift-off process. After fabrication, the chip was diced and polished to reduce the coupling loss. This fabrication process could be further simplified by fabricating the rib with EBL resist instead of silica \cite{yu2019photonic}. Based on this fabrication recipe, we fabricated modulators with several different modulation lengths of 3 mm, 5 mm, 8 mm, and 12.917 mm to test the modulation performance.

\begin{figure}[ht!]
\centering\includegraphics[width=\linewidth]{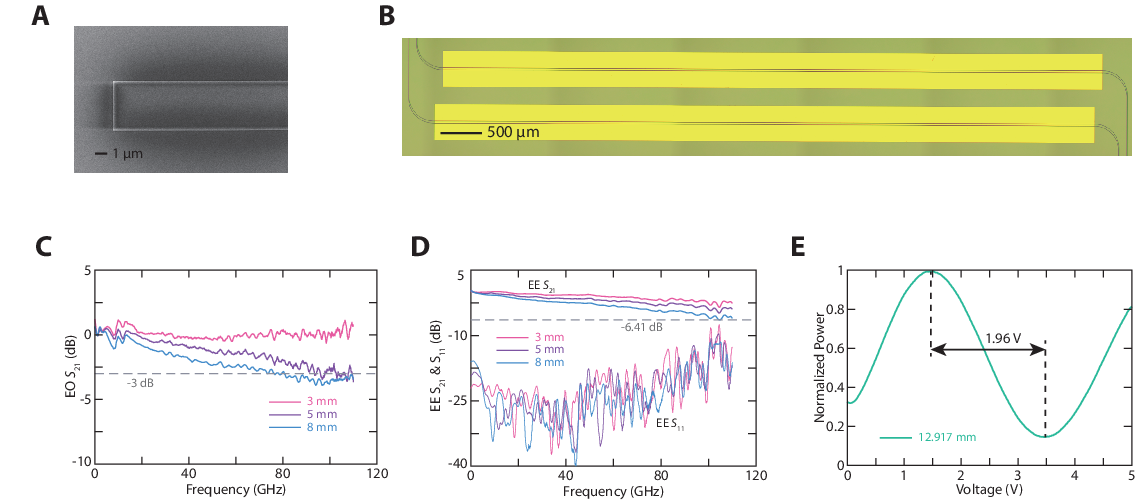}
\caption{Characterization and measurement of silica rib loaded waveguide-based electro-optic modulator. (A) Scanning electron microscope (SEM) image of the silica rib loaded waveguide. (B) Optical microscopy image of the fabricated modulator. (C) Measured electro-optic $S_{21}$. (D) Measured electric $S$ parameters. (E) Measured normalized optical transmission of a 12.917-mm modulator, showing the half-wave voltage at 5 kHz frequency.}
\label{EO}
\end{figure}

We measured the silica rib loaded waveguide-based modulator's electro-optic responses according to a standard measurement process \cite{zhang:2022}. We used a tunable continuous wave (CW) laser to generate CW light near 1550 nm. To excite the TE mode of the rib loaded waveguide, we used a 3-paddle polarization controller to manipulate the polarization of CW light. We used a pair of microwave probes to connect the traveling wave electrodes to an vector network analyzer (VNA) and a 50-$\Omega$ load. When our device is modulated by the microwave signal from VNA, we used an high-speed photodiode to detect the modulated optical signal and send it back to VNA. The responses from the coaxial lines, microwave probes, and the photodiode were calibrated in advance. In this measurement, 3-dB bandwidth is defined as the modulation frequency at which electro-optic response drops for 3 dB compared to that at 1-GHz modulation frequency. As shown in figure \ref{EO}C, the measured electro-optic transmission from VNA shows a flat response above -3 dB of over 110 GHz for the modulator with a modulation length of 3 mm. The 3-dB bandwidths of 5-mm, 8-mm modulators are 94.171 GHz and 80.661 GHz, respectively. Our device's 3-dB bandwidth could be further improved by introducing the micro-structured electrode configuration \cite{kharel2021breaking}.

Using a similar measurement process, we measured the high-speed electric response of the modulator's electrode. We directly measured the modulator's \(S_{11}\) (reflection coefficient) and \(S_{21}\) (transmission coefficient) by connecting VNA and the electrode of modulator via high-speed microwave cables and a pair of high-speed ground-signal-ground probes. As shown in figure \ref{EO}D, for modulators with different modulation lengths, the measured electric transmissions \(S_{21}\) decrease slowly and smoothly as the microwave frequency increases for modulators with different modulation lengths, featuring \(S_{21}\) above -6.41 dB at modulation frequencies up to 110 GHz.

Based on the setup for measuring the electro-optic performance, we measured our device's low-frequency half-wave voltage and extinction ratio. In these measurements, we used a 5-kHz microwave frequency to avoid the photorefractive effect whose response time is at the scale of tens of milliseconds \cite{Jiang2017}. As shown in figure \ref{EO}E, the measured \(V_{\rm \pi}\) of a 12.917-mm modulator is 1.96 V, corresponding to a voltage-length product (\(V_{\rm \pi} L\)) of 2.53 V cm. Among all the available devices, the best measured extinction ratio is 10.73 dB while the best insertion loss is 18.3 dB. The extinction ratio and insertion loss could be further improved by carefully designing the TFLN thickness and the edge coupling structure, respectively.

\begin{figure}[ht!]
\centering\includegraphics[width=\linewidth]{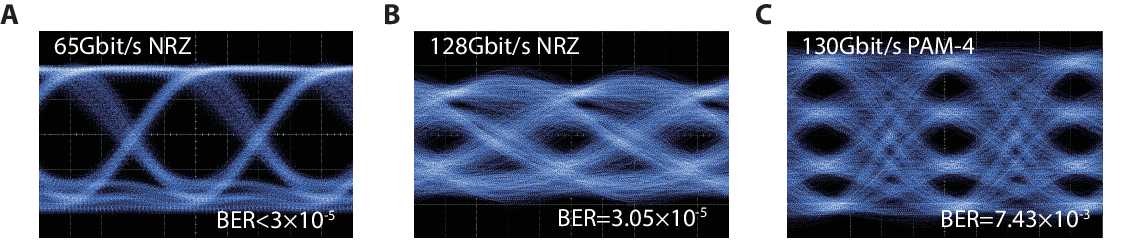}
\caption{Measured eye diagrams of data modulation and transmission with (A) 65 Gbit/s NRZ signal, (B) 128 Gbit/s NRZ signal, and (C) 130 Gbit/s (65 GBaud/s) PAM-4 signal.}
\label{Eye}
\end{figure}

We conducted the data transmission test by using an arbitrary waveform generator (AWG) and a sampling oscilloscope. Nonreturn-to-zero (NRZ) and pulse-amplitude modulation (PAM)-4 signals with symbol transmission rates from 28 GBaud/s to 128 GBaud/s were generated via AWG and were sent into the modulator for digital data modulation. Figure \ref{Eye} show the typical eye diagrams with high symbol transmission rates. For NRZ signal with bit rate up to 128 Gbit/s, the bit error rate (BER) is always lower than \(3.05 \times 10^{-5}\), demonstrating our device's capability in high-speed data communications. We also demonstrated the transmission of PAM-4 signal with bit rate up to 130 Gbit/s (symbol transmission rate of 65 GBaud/s), showing a higher BER of \(7.43 \times 10^{-3}\). These results validate our modulators' potential applications in high-speed data communication. The bit rate and BER of our modulator could be further improved by implementing the in-phase/quadrature modulation scheme or increasing the extinction ratio via an optimal design of edge coupling structure.

\section{Conclusion}

Based on our theoretical analysis of low index rib loaded waveguides for electro-optic modulators, we designed, fabricated, and tested a high-performance integrated lithium niobate modulator without direct etching of lithium niobate. We provided a systematic theoretical study in the fundamental TE mode of low-index rib loaded TFLN waveguide structure. Based on this study, we fabricated and tested an integrated lithium niobate low-index rib loaded waveguide modulator. Compared with other rib loaded waveguide-based TFLN modulators, our modulator features a record-high 3-dB bandwidth of over 110 GHz (Table~\ref{table}), a symbol transmission rate up to 128 GBaud/s with BER down to \(3.05 \times 10^{-5}\), and a voltage-length product of 2.53 V cm. The performances of this design are comparable to those of etched TFLN modulators. Our design provides a solution to the mass production of high-performance integrated TFLN modulators at low cost.

\begin{table}[htbp]
  \centering
  \caption{Performance comparison of rib loaded waveguide-based TFLN modulators}
    \begin{tabular}{ccccc}
    \toprule
    \textbf{Reference} & \textbf{Year} & \textbf{3dB Bandwidth (GHz)} & \textbf{\(V_{\rm \pi} L\) (V cm)} & \textbf{Rib material} \\
    \midrule
    \cite{rabiei2013heterogeneous} & 2013  & /     & 4     & Ta\(_2\)O\(_5\) \\
    \cite{cao2014hybrid} & 2014  & 3.5   & 1.6   & \(\alpha\)-Si \\
    \cite{rao2015heterogeneous} & 2015  & 1     & 3.8   & Chalcogenide glass \\
    \cite{jin2015linbo} & 2015  & 8     & 3     & SiN\(_x\) \\
    \cite{rao2016high} & 2016  & 33    & 6.5   & SiN\(_x\) \\
    \cite{ahmed2020subvolt} & 2020  & /     & 2.11  & SiN\(_x\) \\
    \cite{ahmed2020high} & 2020  & 29    & 3.12  & Si\(_3\)N\(_4\) \\
    \cite{boynton2020heterogeneously} & 2020  & 30.6  & 6.7   & SiN\(_x\) \\
    \cite{huang2021high} & 2021  & > 40   & 3.1   & SiN\(_x\) \\
    \cite{zhang2021high} & 2021  & 30    & 2.18  & SiN\(_x\) \\
    \cite{nelan2022ultra} & 2022  & 30    & 3.3   & SiN\(_x\) \\
    \textbf{Our work} & \textbf{2023} & \textbf{> 110} & \textbf{2.53} & \textbf{SiO\(_2\)} \\
    \bottomrule
    \end{tabular}%
  \label{table}%
\end{table}%

\begin{backmatter}
\bmsection{Funding}
This work was supported by National Key Research and Development Program of China (2021YFA-1401000); National Natural Science Foundation of China (62075114); Beijing Municipal Natural Science Foundation (4212050, Z220008); Zhuhai Industry University Research Collaboration Project (ZH-22017001210108PWC). This work was supported by the Center of High Performance Computing, Tsinghua University. 

\bmsection{Acknowledgments}
The authors thank Rongjin Zhuang, Jinze He and Tian Dong for the helpful discussion.

\bmsection{Disclosures}
The authors declare no conflicts of interest.

\bmsection{Data availability}
Data underlying the results presented in this paper are not publicly available at this time but may be obtained from the authors upon reasonable request.

\end{backmatter}

\end{document}